\documentclass[11pt,onecolumn,amssymb,nofootinbib]{revtex4}
\usepackage{amsmath, amsthm, amscd, amssymb}
\usepackage{bm}
\usepackage{bbm}

\begin{document}

\title{\bf Collapse models with non-white noises}
\author{Stephen L. Adler}
\email{adler@ias.edu} \affiliation{Institute for Advanced Study,
Einstein Drive, Princeton, NJ 08540, USA.}
\author{Angelo Bassi}
\email{bassi@ts.infn.it, bassi@mathematik.uni-muenchen.de}
\address{Dipartimento di Fisica Teorica,
Universit\`a di Trieste, Strada Costiera 11, 34014 Trieste, Italy.
\\  Mathematisches Institut der L.M.U., Theresienstr. 39, 80333
M\"unchen, Germany. \\ Istituto Nazionale di Fisica Nucleare,
Sezione di Trieste, Strada Costiera 11, 34014 Trieste, Italy.}
\begin{abstract}
We set up a general formalism for models of spontaneous wave
function collapse with dynamics represented by a stochastic
differential equation driven by general Gaussian noises, not
necessarily white in time. In particular, we show that the
non-Schr\"odinger terms of the equation induce the collapse of the
wave function to one of the common eigenstates of the collapsing
operators, and that the collapse occurs with the correct quantum
probabilities. We also develop a perturbation expansion of the
solution of the equation with respect to the parameter which sets
the strength of the collapse process; such an approximation allows
one  to compute the leading order terms for the deviations of the
predictions of collapse models with respect to those of standard
quantum mechanics. This analysis shows that to leading order, the
``imaginary noise'' trick can be used for non-white Gaussian noise.
\end{abstract}
\maketitle

\section{Introduction}
\label{sec:one}

Models of spontaneous wave function collapse~\cite{cm} provide a
simple consistent resolution to the measurement problem of quantum
mechanics~\cite{mp}, and at the same time provide precise
indications for experiments which are more likely to detect
possible violations of quantum linearity~\cite{rev}: these include
e.g. fullerene diffraction experiments, decay of supercurrents,
excitation of bound atomic and nuclear systems and several
cosmological observations. The dynamics  is represented by a
stochastic Schr\"odinger equation of the form:
\begin{equation} \label{eq:cme}
d|\psi_{t}\rangle \; = \; \left[ - \frac{i}{\hbar} \, H \, dt \; +
\; \sqrt{\gamma}\, \sum_{i = 1}^{N} (A_{i} - \langle A_{i}
\rangle_{t})\, d W_{i,\,t} \; - \; \frac{\gamma}{2} \sum_{i=1}^{N}
(A_{i} - \langle A_{i} \rangle_{t})^2\, dt \right]
|\psi_{t}\rangle,
\end{equation}
where $H$ is the standard quantum Hamiltonian of the system,
$A_{i}$ are a set of commuting self-adjoint operators to whose
eigenstates the wave function is driven during the collapse
process, $W_{i,\,t}$ are $N$ independent standard Wiener processes
defined on a probability space $(\Omega, {\mathcal F}, {\mathbb
P})$, the average $\langle A_{i} \rangle_{t} \equiv \langle
\psi_{t} | A_{i} | \psi_{t}\rangle$ is the standard quantum
expectation of $A_{i}$ and $\gamma$ is a positive constant which
sets the strength of the collapse process.

The several collapse models which have been so far proposed differ
from each other basically only by the choice of the localizing
operators: in GRW-type models~\cite{pos}, the set $\{ A_{i} \}$
corresponds to the set of position operators of the constituents
of the given physical system, or some function of them;
dissipative effects can be included by taking $A_{i}$ to be a
function of both the position and the momentum operator of a
particle~\cite{dis}, the resulting operator being non-Hermitian;
in the CSL model for identical particles~\cite{csl}, the index $i$
is replaced by the space coordinate ${\bf x}$ and $A({\bf x})$
bacomes a function of the density number operator
$a^{\dagger}({\bf x})a({\bf x})$; in energy driven
models~\cite{em} there appears only one operator $A$, which is
identified with the Hamiltonian $H$. Finally, reduction models
related to gravitational effects~\cite{pen} can also be cast in
the form~\eqref{eq:cme}, as shown in~\cite{di}.

The reduction properties of Eq.~\eqref{eq:cme} can be easily
verified by computing the time evolution of the variance $V_{A}(t)
\equiv \langle A^2 \rangle_{t} - \langle A \rangle_{t}^2$, of an
operator $A$ which commutes with all the operators $A_{i}$; as shown
e.g. in~\cite{em}, by using standard It\^o calculus rules, and by
setting $H = 0$, one gets for the average value ${\mathbb
E}_{\mathbb P} [ V_{A}(t) ]$ the following equation:
\begin{equation} \label{eq:var}
{\mathbb E}_{\mathbb P} [ V_{A}(t) ] \quad = \quad V_{A}(0) \; -
\; 4 \gamma\, \sum_{i=1}^{N} \int_{0}^{t} ds\, {\mathbb
E}_{\mathbb P} [ C_{A,A_{i}}^2(s) ],
\end{equation}
with $C_{A,A_{i}}(t) \equiv \langle (A - \langle A
\rangle_{t})(A_{i} - \langle A_{i} \rangle_{t}) \rangle_{t}$. Since
the integrand on the right hand side is a non-negative quantity, the
above relation, when applied to any operator $A_{i}$, implies that,
for large times, the variance $V_{A_{i}}(t)$ converges to 0 for any
realization of the noise, with the possible exception only of a
subset of $\Omega$ of measure 0; this means that any initial state
$|\psi_{0}\rangle$ converges asymptotically, with probability 1, to
one of the common eigenstates of the operators $A_{i}$. When $H \neq
0$ and moreover it does not commute with the other operators
$A_{i}$, Eq.~\eqref{eq:cme} induces only an approximate collapse,
the degree of  approximation depending on the relative strength of
the Schr\"odinger term and of the collapse terms which define the
equation.

A very useful mathematical property of Eq.~\eqref{eq:cme} is that
its physical predictions concerning the outcomes of measurements
are, in terms of statistical expectations, invariant under a phase
change in the noise. As a matter of fact, let us consider the
following class of stochastic Schr\"odinger equations:
\begin{equation} \label{eq:cpe}
d|\psi_{t}\rangle = \left[ - \frac{i}{\hbar} H dt + \sqrt{\gamma}
\sum_{i = 1}^{N} (\xi A_{i} - \xi_{\text{\tiny R}} \langle A_{i}
\rangle_{t}) d W_{i,\,t} - \frac{\gamma}{2} \sum_{i=1}^{N}
(|\xi|^2 A_{i}^2 - 2\xi\xi_{\text{\tiny R}} A_{i} \langle A_{i}
\rangle_{t} + \xi_{\text{\tiny R}}^2 \langle A_{i} \rangle_{t}^2)
dt \right] |\psi_{t}\rangle,
\end{equation}
where $\xi = \xi_{\text{\tiny R}} + i \xi_{\text{\tiny I}} $ is a
constant complex factor; of course, when $\xi = 1$ we recover our
original collapse equation. An easy application of It\^o calculus
leads to the following equation for the density matrix $\rho(t) =
{\mathbb E}_{\mathbb P} [ |\psi_{t}\rangle \langle\psi_{t}| ]$:
\begin{equation} \label{eq:lind}
\frac{d}{dt}\, \rho(t) \; = \; - \frac{i}{\hbar}\, [ H, \rho(t) ] \;
+ \; \frac{\gamma}{2} \, |\xi|^2 \, \sum_{i = 1}^{N} \left( 2\,
A_{i}\,\rho(t)\, A_{i} - \{ A_{i}^2, \rho(t) \} \right),
\end{equation}
which is of the Lindblad type and has the remarkable property that
it depends only on the square modulus of $\xi$. Since, within
collapse models, the statistics of the outcome of
experiments~\cite{obs} can be expressed by the averages ${\mathbb
E}_{\mathbb P} [ \langle \psi_{t} | O | \psi_{t} \rangle ] \equiv
\text{Tr} [ \rho(t) O ]$, where $O$ is a self-adjoint operator, we
see that in order to compute experimental predictions, one can use
in place of Eq.~\eqref{eq:cme} any stochastic equation of the
type~\eqref{eq:cpe} which satisfies the constraint $|\xi| = 1$; in
most cases, it is convenient to choose the equation corresponding to
$\xi = i$, since it is linear, thus much easier to solve~\cite{pp}.
Of course this does not mean that all equations of the
form~\eqref{eq:cpe}, having the same value of $|\xi|$, are
equivalent; on the contrary, in general they generate completely
different evolutions for the wave function. For example, when $\xi =
1$, as we have seen, the corresponding equation induces collapse of
the wave function, since the property~\eqref{eq:var} for the
variance holds true; while on the other hand with $\xi = i$, the
corresponding equation is linear, thus the wave function does not
collapse in this case.  This notwithstanding, all quantities of the
form ${\mathbb E}_{\mathbb P} [ \langle \psi_{t} | O | \psi_{t}
\rangle ]$ turn out to be the same for the two equations, and for
similar ones with complex $\xi$ of modulus unity.

The aim of this paper is to generalize Eq.~\eqref{eq:cme} in order
to include also non-white, Gaussian, stochastic processes. Some
results have already appeared in the
literature~\cite{col,bg,srt,pnw}, but a general analysis is still
lacking, mainly because the new dynamics is not Markovian and thus
is more difficult to describe mathematically. There are two main
reasons why one should consider collapse models driven by
non-white noises. First of all, it is important to understand how
the collapse mechanism and other physical properties, such as the
time evolution of the mean energy, depend on the type of  noise
driving the collapse of the wave function.  Since these properties
are directly connected to physical predictions which differ from
those given by standard quantum mechanics, such differences, and
thus the possibility of experimentally testing collapse models,
could significantly change depending on the type of noise entering
the collapse equation. The second reason for such an analysis is
that a non-white noise, unlike a Wiener process, can be identified
with a  physical field; accordingly, one can try to connect the
collapse mechanism to some other physical process occurring in
Nature, possibly having a cosmological origin; we will come back
to this point in the final section.

This paper contains two main sections, which set up the general
formalism for non-Markovian collapse models by following two
different paths: in Sec.~\ref{sec:two} we follow the same argument
used in~\cite{csl} to derive, from a generic diffusion process in
Hilbert space, Eq.~\eqref{eq:cme} as the correct collapse
equation; in Sec.~\ref{sec:three} instead we follow the strategy
used in~\cite{ad} to obtain the same equation from general
requirements on the dynamics of the density matrix.  We will prove
the two following main results:
\begin{enumerate}
\item
We will show that, if one neglects the quantum Hamiltonian $H$, the
dynamics leads to the collapse of the wave function to one of the
common eigenstates of the localizing operators, with the correct
quantum probabilities.

\item We will develop a perturbation expansion of the solution of
the equation with respect to the coupling constant
$\sqrt{\gamma}$; when applied to experimental predictions on
microscopic systems, it provides the leading order term for the
deviation of such predictions from those given by standard quantum
mechanics.
\end{enumerate}
Concerning this second result, we will also show that, at least to
order $\gamma$, the equation for the statistical operator depends
only on the absolute value of $\xi$, precisely as discussed before
for the white-noise case; this means that, to this order, one can
employ the useful trick of replacing the real-noise ($\xi = 1$),
non-linear collapsing equation with an imaginary-noise ($\xi =
i$), linear non-collapsing equation, thus  considerably
simplifying calculations.

Throughout the calculations, one has to make sure that, at any
stage, one recovers the correct white-noise limit; however, one
has to keep in mind that the white-noise limit of the non-white
collapsing equation will not be given by~\eqref{eq:cme}
or~\eqref{eq:cpe}, since these are It\^o equations; instead, as
explained e.g. in~\cite{arn}, an equation containing non-white
noises reduces, in the white-noise limit, to a Stratonovich
equation. Accordingly, we write the Stratonovich equation
corresponding to the It\^o equation~\eqref{eq:cpe}, which
is~\cite{csl}:
\begin{eqnarray} \label{eq:str}
\frac {d|\psi_{t}\rangle }{dt}& = & \left[ - \frac{i}{\hbar} H  +
\sqrt{\gamma} \sum_{i = 1}^{N} (\xi A_{i} - \xi_{\text{\tiny R}}
\langle A_{i} \rangle_{t}) w_i(t) \right. \nonumber \\ & &
\qquad\qquad\left. - \gamma\xi_{\text{\tiny R}} \sum_{i=1}^{N}
(\xi A_{i}^2 - 2\xi A_{i} \langle A_{i} \rangle_{t} -
\xi_{\text{\tiny R}} \langle A_{i}^2 \rangle_{t} +
2\xi_{\text{\tiny R}} \langle A_{i} \rangle_{t}^2)  \right]
|\psi_{t}\rangle\cr
 & = & \left[ - \frac{i}{\hbar} H  + \sqrt{\gamma} \sum_{i =
1}^{N} (\xi A_{i} - \xi_{\text{\tiny R}}
\langle A_{i} \rangle_{t}) (w_i(t)+ 2\sqrt{\gamma}\xi_{\text{\tiny R}} \langle A_{i} \rangle_{t}) \right. \nonumber \\
& & \qquad\qquad\left. - \gamma\xi_{\text{\tiny R}} \sum_{i=1}^{N}
(\xi A_{i}^2  - \xi_{\text{\tiny R}} \langle A_{i}^2 \rangle_{t})
\right] |\psi_{t}\rangle.
\end{eqnarray}
In the second expression, we have written the third term in square
brackets in a form that corresponds to the white noise limit of
Eq.~\eqref{eq:colcomcon} below, with the remainder of the order
$\gamma$ part playing the role of a shift in the mean value of the
noise, corresponding to a change of measure, as discussed in Sec.
IV below.  Of course, whether one uses Eq.~\eqref{eq:cpe}
or~\eqref{eq:str}, the corresponding equation for the statistical
operator $\rho(t)$, which determines the evolution of statistical
ensembles of states, will always be given by~\eqref{eq:lind}.

\section{Linear and nonlinear collapse equation}
\label{sec:two}

Following the path outlined in~\cite{csl} to construct the
continuous generalization of the GRW model in terms of an It\^o
stochastic differential equation of the type~\eqref{eq:cme}, let
us consider a diffusion process for the wave function in Hilbert
space having the form:
\begin{equation} \label{eq:cm1}
\frac{d|\phi(t)\rangle}{dt} \quad = \quad \left[ -\frac{i}{\hbar}
H\; +\; \sqrt{\gamma}\, \xi\, \sum_{i=1}^{N} A_{i}w_{i}(t) \; + \; O
\right] |\phi(t)\rangle,
\end{equation}
where, as before, $H$ is the standard quantum Hamiltonian of the
system, $A_{i}$ are commuting self-adjoint operators, $\gamma$ is a
positive coupling constant, $\xi = \xi_{\text{\tiny R}} + i
\xi_{\text{\tiny I}} $ is a constant complex factor, while $O$ is a
linear operator yet to be defined\footnote{As we shall see, the
operator $O$ will not be a standard linear operator, since it will
act on a vector also through its dependence on the noises
$w_{i}(t)$, by means of functional derivatives. In this respect, our
approach differs from the standard one based on It\^o's diffusion
equations in Hilbert spaces.}.  The noises $w_{i}(t)$ are real
Gaussian random processes defined on a probability space $(\Omega,
{\mathcal F}, {\mathbb Q})$ whose mean and correlation functions
are, respectively:
\begin{equation} \label{eq:cm2}
{\mathbb E}_{\mathbb Q} [ w_{i}(t) ] = 0, \qquad {\mathbb
E}_{\mathbb Q} [ w_{i}(t_{1}) w_{j}(t_{2}) ] = D_{ij}(t_{1},t_{2}).
\end{equation}
When $\xi_{\text{\tiny R}} \neq 0$, which is the case for collapse
models, Eq.~\eqref{eq:cm1} does not preserve the norm of the wave
function; therefore we introduce the normalized vector:
\begin{equation} \label{eq:cm3}
|\psi(t)\rangle \; = \; \frac{|\phi(t)\rangle}{\| |\phi(t)\rangle
\|},
\end{equation}
(assuming of course that the norm of $|\phi(t)\rangle$ does not
vanish) which we take as the physical vector describing the random
state of the system at time $t$.

The measure ${\mathbb Q}$ previously introduced is not the correct
physical probability since it does not lead to a collapse that
respects  the Born probability rule; the right physical
probability, which we shall call ${\mathbb P}$, is defined as
follows:
\begin{equation} \label{eq:cm4}
{\mathbb P}[F] \; = \; {\mathbb E}_{\mathbb Q} [ 1_{F} \langle
\phi(t) | \phi(t) \rangle ] \qquad \forall \,\, F \, \in \,
{\mathcal F},
\end{equation}
where $1_{F}$ is the indicator function associated to the measurable
subset $F$ of $\Omega$. This definition corresponds to the
assumption of the GRW model, according to which a collapse (in
space) is more likely to occur where the wave function is larger, as
postulated by the Born probability rule.

To summarize, an initial state $|\psi(0)\rangle$ is driven by the
stochastic dynamics into an ensemble of states $|\psi(t)\rangle$ of
the form~\eqref{eq:cm3}, where $|\phi(t)\rangle$ solves
Eq.~\eqref{eq:cm1}; the distribution of states within the ensemble
is given by the probability ${\mathbb P}$ defined in~\eqref{eq:cm4}.

Note that the definition~\eqref{eq:cm4} of ${\mathbb P}$ not only
matches with the Born probability rule, but is also necessary in
order to prevent the possibility of using the collapse mechanism to
send information at a speed faster than the speed of light. As shown
in~\cite{gisin}, when the dynamics of a statistical operator
$\rho(t)$ is nonlinear, it is in general possible to send
faster-than-light signals which can be used by two space-like
separated observers to communicate with each other; this possibility
is instead forbidden when the evolution is linear. In our case,
according to the previous assumptions, $\rho(t)$ is defined as
follows:
\begin{equation} \label{eq:cm5}
\rho(t) \; = \; {\mathbb E}_{\mathbb P}\, [ | \psi(t) \rangle
\langle \psi(t) | ];
\end{equation}
but according to Eqs.~\eqref{eq:cm3} and~\eqref{eq:cm4} we have the
mathematical equality:
\begin{equation} \label{eq:cm6}
{\mathbb E}_{\mathbb P}\, [ | \psi(t) \rangle \langle \psi(t) | ] \;
= \; {\mathbb E}_{\mathbb Q}\, [ | \phi(t) \rangle \langle \phi(t) |
],
\end{equation}
and since $| \phi(t) \rangle$ solves the linear
equation~\eqref{eq:cm1}, it follows that the operator which maps
$\rho(t)$ to $\rho(t+dt)$ is linear, thus not allowing
faster-than-light signalling of the type just discussed.

We still need to prove that Eq.~\eqref{eq:cm4} correctly defines a
probability measure; it is easy to check that all properties are
satisfied expect for the normalization condition ${\mathbb
P}[\Omega] = {\mathbb E}_{\mathbb Q}[ \langle \phi_{t} | \phi_{t}
\rangle ] = 1$, which in general is not fulfilled unless the
operator $O$ takes a particular form. In fact, by using the
Furutsu-Novikov formula~\cite{fnf}:
\begin{equation} \label{fn}
{\mathbb E}_{\mathbb Q} [ F[\{w(t)\}]\,w_{i}(t) ] \quad = \quad
\sum_{j=1}^{N} \int_{{0}}^{t} ds \, D_{ij}(t,s) {\mathbb
E}_{\mathbb Q} \left[ \frac{\delta F[\{w(t)\}]}{\delta w_{j}(s)}
\right]
\end{equation}
which holds for a generic functional $F[\{w(t)\}]$ of the Gaussian
noises $w_{i}(t)$, $i = 1, \ldots N$ satisfying~\eqref{eq:cm2},
and computed from initial time $0$ to time $t$, one can
immediately prove that:
\begin{equation} \label{eq:cm6bis}
\frac{d}{dt}\, {\mathbb E}_{\mathbb Q}[ \langle \phi_{t} |
\phi_{t} \rangle ] \; = \; 0 \qquad \text{if:} \qquad O \; = \;
-2\sqrt{\gamma}\, \xi_{\text{\tiny R}} \sum_{i,j=1}^{N} A_{i}
\int_{{0}}^{t}ds\, D_{ij}(t,s) \frac{\delta}{\delta w_{j}(s)}.
\end{equation}
Accordingly, the linear equation which, together with
Eq.~\eqref{eq:cm4}, induces collapse of the wave function with the
correct quantum probabilities and, at the same time,  does not
allow one to use the collapse process to send signals at
faster-than-light speed, is~\cite{bg}:
\begin{equation} \label{eq:col}
\frac{d|\phi(t)\rangle}{dt} \quad = \quad \left[ -\frac{i}{\hbar}
H\; +\; \sqrt{\gamma}\, \xi\, \sum_{i=1}^{N} A_{i}w_{i}(t) \; - \;
2\sqrt{\gamma}\, \xi_{\text{\tiny R}} \sum_{i,j=1}^{N} A_{i}
\int_{{0}}^{t}ds\, D_{ij}(t,s) \frac{\delta}{\delta w_{j}(s)}
\right] |\phi(t)\rangle.
\end{equation}
As foreseen, this equation is non-Markovian and for this reason is
highly non-trivial, since the future evolution, which involves the
whole past, depends on the combined effect of the standard
Hamiltonian $H$ and the collapsing operators $A_{i}$.  This
dynamics  is not easy to unfold if these operators do not commute
among themselves, as is usually the case.

The equation for the normalized vector $|\psi(t)\rangle$ does not
have a closed form, unless the functional derivative of
$|\phi(t)\rangle$ can be explicitly computed; as we shall se in the
next sections, this happens when the Hamiltonian $H$ is neglected
(or when it commutes with the operators $A_{i}$), and when one
writes the evolution as a perturbation expansion with respect to the
relevant parameters.

Before moving on, we make a few comments about the change of measure
as defined in~\eqref{eq:cm4}. In the white noise case one can prove,
under suitable hypotheses on the operators $H$ and $A_{i}$ (see
e.g.~\cite{hol}), that $\langle \phi(t) | \phi(t) \rangle$ is
martingale with ${\mathbb Q}$-mean equal to 1; this ensures that one
can consistently use $\langle \phi(t) | \phi(t) \rangle$ as a
Radon-Nikodym derivative of a new probability measure ${\mathbb P}$
with respect to ${\mathbb Q}$, as we have assumed more heuristically
in the previous paragraphs. Here we will not attempt to prove that
$\langle \phi(t) | \phi(t) \rangle$ satisfies the required
properties also in the more general case of non-white Gaussian
processes, leaving the analysis of the conditions under which this
is true to future research. Secondly, Girsanov's theorem provides,
in the white noise case, a connection between Wiener processes with
respect to the measure ${\mathbb Q}$ and Wiener processes with
respect to the transformed measure ${\mathbb P}$. It would be
interesting to see whether a similar theorem can be proved also in
the non-white noise case, and whether ${\mathbb Q}$-Gaussian
processes can be connected to ${\mathbb P}$-Gaussian processes; we
will come back on this point in Sec. IV.

\subsection{Collapse of the state vector}

We now show that, when the standard quantum Hamiltonian $H$ is set
to 0, the dynamics induces the collapse of the state vector
$|\psi(t)\rangle$ to one of the common eigenstates of the operators
$A_{i}$. As shown in~\cite{bg}, if $H$ is neglected so that all
operators entering the equation commute among themselves, then the
functional derivative can be explicitly computed and
Eq.~\eqref{eq:col} reduces to:
\begin{equation} \label{eq:colcom}
\frac{d|\phi(t)\rangle}{dt} \quad = \quad \left[ \sqrt{\gamma}\,
\xi\, \sum_{i=1}^{N} A_{i}w_{i}(t) \; - \; 2\gamma\, \xi
\xi_{\text{\tiny R}} \sum_{i,j=1}^{N} A_{i} A_{j} F_{ij}(t) \right]
|\phi(t)\rangle,
\end{equation}
where we have defined:
\begin{equation}\label{eq:fdeff}
F_{ij}(t) \; = \; \int_0^{t} ds \, D_{ij}(t,s).
\end{equation}
That Eq.~\eqref{eq:colcom} is equivalent to Eq.~\eqref{eq:col} in
the limit $H = 0$ can  easily be seen by integration of
Eq.~\eqref{eq:colcom}, which is trivial since all operators
commute, from which one obtains the relation:
\begin{equation}\label{eq:phiderivvv}
\frac{\delta}{\delta w_{j}(s)}\, |\phi(t)\rangle \; = \;
\sqrt{\gamma}\, \xi\, A_{j} |\phi(t)\rangle, \qquad s \leq t,
\end{equation}
a relation which will often be used in the following calculations.

The equation for the normalized vector $|\psi(t)\rangle$ can now be
directly computed from the definition~\eqref{eq:cm3}:
\begin{equation} \label{eq:colcomcon}
\frac{d|\psi(t)\rangle}{dt} =  \left[ \sqrt{\gamma} \sum_{i=1}^{N}
(\xi A_{i} - \xi_{\text{\tiny R}} \langle A_{i} \rangle_{t})
w_{i}(t) - 2\gamma \xi_{\text{\tiny R}}\! \sum_{i,j=1}^{N} (\xi
A_{i} A_{j} - \xi_{\text{\tiny R}}  \langle A_{i} A_{j}
\rangle_{t}  ) F_{ij}(t) \right] |\psi(t)\rangle,
\end{equation}
with the expectations $\langle ...\rangle_{t}$ computed in the state
$|\psi(t)\rangle$, that is $\langle O \rangle_t\equiv \langle
\psi(t)|O|\psi(t)\rangle$.

 The equation for the statistical
operator $\rho(t)$ can now be computed either from
Eq.~\eqref{eq:colcomcon} through the definition of
Eq.~\eqref{eq:cm5} or from Eq.~\eqref{eq:colcom} through the
equivalence of Eq.~\eqref{eq:cm6}; in both cases one gets:
\begin{equation} \label{eq:lindnw}
\frac{d}{dt}\, \rho(t) \; = \;  \gamma \, |\xi|^2 \, \sum_{i,j =
1}^{N} \left( A_{i}\,\rho(t)\, A_{j} + A_{j}\,\rho(t)\, A_{i} -
A_{i}A_{j} \rho(t) - \rho(t) A_{j}A_{i} \right) F_{ij}(t),
\end{equation}
which correctly reduces to~\eqref{eq:lind} in the white-noise
limit.\footnote{Note that in the white noise limit, through
Eq.~\eqref{eq:fdeff} one encounters the integral of a delta function
at the endpoint of an interval, which is $\int_0^t ds
\delta(t-s)=1/2$, since $\delta(t)=\delta(-t)$ and
$\int_{-\infty}^{\infty}dt\delta(t)=1$.  This enters both in comparing the
white noise limit of Eq.~\eqref{eq:lindnw} to Eq.~\eqref{eq:lind},
and the white noise limit of Eq.~\eqref{eq:colcomcon} to
Eq.~\eqref{eq:str}.}

We have now all the necessary formulas to compute the time evolution
of quantities such as ${\mathbb E}_{\mathbb P} [ \langle A^{n} \rangle_{t} ]$ and ${\mathbb E}_{\mathbb P} [ \langle A
\rangle_{t}^{n} ]$, where $A$ is a self-adjoint operator commuting
with all the operators $A_{i}$. Because of the relation ${\mathbb
E}_{\mathbb P} [ \langle A^{n} \rangle_{t} ] = \text{Tr} [ A^{n}
\rho(t) ]$, which is a consequence of Eq.~\eqref{eq:cm5}, and
because of the trace-preserving structure of Eq.~\eqref{eq:lindnw},
one immediately has:
\begin{equation} \label{eq:anm}
\frac{d}{dt}\, {\mathbb E}_{\mathbb P} [ \langle A^{n} \rangle_{t} ]
\; = \; 0.
\end{equation}
By using the change of measure~\eqref{eq:cm4} and through a direct
calculation, one finds for ${\mathbb E}_{\mathbb P} [ \langle A
\rangle_{t}^{n} ]$ instead that:
\begin{eqnarray} \label{eq:dfg}
\frac{d}{dt}\, {\mathbb E}_{\mathbb P} [ \langle A \rangle_{t}^{n}
] & = &2 \sqrt{\gamma}\, \xi_{\text{\tiny R}} {\mathbb E}_{\mathbb
P} [\langle A \rangle_{t}^{n-1} ( n \langle A X \rangle_{t} -
(n-1) \langle A \rangle_{t} \langle X
\rangle_{t}) ], \nonumber \\
X & = &\sum_{i=1}^N A_i w_i(t)-2 \sqrt{\gamma}\, \xi_{\text{\tiny
R}} \sum_{i,j=1}^N A_iA_jF_{ij}(t).
\end{eqnarray}
The terms proportional to $w_{i}(t)$ can be rewritten by using the
Furutsu-Novikov formula, together with the equality:
\begin{equation}
\frac{\delta \langle O \rangle_{t}}{\delta w_{j}(s)} \; = \;
\sqrt{\gamma} [ \xi^{\star} \langle A_{j} O \rangle_{t} + \xi
\langle O A_{j} \rangle_{t} -2\xi_{\text{\tiny R}} \langle O
\rangle_{t}\langle A_{j} \rangle_{t}],
\end{equation}
which is valid for any operator $O$, and which can be directly
proved from the definition $\langle O \rangle_{t} \equiv \langle
\phi(t) | O | \phi(t) \rangle / \langle \phi(t) | \phi(t)
\rangle$, together with Eq.~\eqref{eq:phiderivvv}.   After a
rather lengthy calculation, one can prove that Eq.~\eqref{eq:dfg}
simplifies to:
\begin{equation} \label{eq:amn}
\frac{d}{dt}\, {\mathbb E}_{\mathbb P} [ \langle A\rangle_{t}^{n}
] \; = \; 4  n(n-1) \gamma\, \xi_{\text{\tiny R}}^2 \sum_{i.j}
{\mathbb E}_{\mathbb P} [ \langle A \rangle^{n-2}_{t} \langle
(A_{i} - \langle A_{i} \rangle_{t}) A \rangle_{t} \langle (A_{j} -
\langle A_{j} \rangle_{t}) A \rangle_{t}] \, F_{ij}(t).
\end{equation}

We now apply Eqs.~\eqref{eq:anm} and~\eqref{eq:amn} to compute the
time evolution of the variance $V_{A}(t) = \langle A^2 \rangle_{t} -
\langle A \rangle_{t}^2$ of the operator $A$; we obtain:
\begin{equation} \label{eq:redu}
{\mathbb E}_{\mathbb P} [ V_{A}(t) ] \quad = \quad V_{A}(0) \; -
\; 8\, \xi_{\text{\tiny R}}^2\, \gamma \,\sum_{i.j} \int_0^{t} ds
\, {\mathbb E}_{\mathbb P} [ \langle (A_{i} - \langle A_{i}
\rangle_{s}) A \rangle_{s} \langle (A_{j} - \langle A_{j}
\rangle_{s}) A \rangle_{s}] \, F_{ij}(s).
\end{equation}
Now, the same argument used in Eq.~\eqref{eq:var}  to prove the
reduction of Eq.~\eqref{eq:cme} holds true: when the matrix
$F_{ij}(s)$ is positive definite in the limit as $t \to \infty$,
Eq.~\eqref{eq:redu} is consistent if only if, for large times,
$\langle (A_{i} - \langle A_{i} \rangle_{t}) A \rangle_{t}$ goes
to zero for any $i$ almost surely (a.s.) (i.e.,  except on a
subset of $\Omega$ of realizations of the noise of ${\mathbb
P}$-measure 0); in particular, if we take $A$ equal to any one of
the operators $A_{i}$ we have:
\begin{equation}
\lim_{t \rightarrow \infty} [\langle A_{i}^2 \rangle_{t} - \langle
A_{i} \rangle_{t}^2 ] \; = \; \lim_{t \rightarrow \infty}
V_{A_{i}}(t) \; = \; 0 \qquad \text{a.s.} \quad \forall \,\, i,
\end{equation}
which is the desired result. Moreover, due to Eq.~\eqref{eq:anm},
the average value of $\langle P_{a_{i}} \rangle_{t}$ remains
constant in time, with  $P_{a_{i}} $ the  projector on any
eigenspace of $A_i$ with eigenvalue $a_i$, which means that the
collapse occurs with the correct quantum probabilities.

\subsection{Perturbation expansion to order $\gamma$}

The approximation used in the previous subsection, which consisted
in neglecting the quantum Hamiltonian $H$, is useful when the
system under study is macroscopic, since in this case the effect
of the collapsing terms is typically much stronger than that of
$H$: this is precisely the reason why collapse models ensure the
localization of the wave function at the macroscopic level. For
microscopic systems, on the other hand, such an approximation is
no longer valid; just the reverse from  the macroscopic case, at
the microscopic level the effect of the collapsing terms
represents typically only a small perturbation on the standard
quantum evolution: for this reason collapse models agree very well
with standard quantum mechanical predictions. It then becomes
meaningful, for micro-systems, to perform a perturbation expansion
of the evolution of the state vector with respect to the parameter
$\sqrt{\gamma}$, in order to compute the leading terms
representing the deviations of the predictions of collapse models
from those given by standard quantum mechanics. To this end, let
us introduce the interaction picture operators and states:
\begin{equation}
A_{i}(t) \; = \; U^{\dagger}(t) A_{i} U(t), \qquad
|\phi^{\text{\tiny I}}(t) \rangle \; = \; U^{\dagger}(t) |\phi(t)
\rangle, \qquad U(t)=\exp(-\frac{i}{\hbar} H t);
\end{equation}
Eq.~\eqref{eq:col} then becomes:
\begin{equation} \label{eq:colint}
\frac{d|\phi^{\text{\tiny I}}(t)\rangle}{dt} \quad = \quad \left[
\sqrt{\gamma}\, \xi\, \sum_{i=1}^{N} A_{i}(t) w_{i}(t) \; - \;
2\sqrt{\gamma}\, \xi_{\text{\tiny R}} \sum_{i,j=1}^{N} A_{i}(t)
\int_0^{t}ds\, D_{ij}(t,s) \frac{\delta}{\delta w_{j}(s)} \right]
|\phi^{\text{\tiny I}}(t)\rangle.
\end{equation}
The perturbation expansion  of $|\phi^{\text{\tiny I}}(t)\rangle$
with respect to the parameter $\sqrt{\gamma}$ reads:
\begin{equation}
|\phi^{\text{\tiny I}}(t)\rangle \; = \; |\phi^{\text{\tiny
I}}_{0}(t)\rangle \, + \, \sqrt{\gamma}\, |\phi^{\text{\tiny
I}}_{1}(t)\rangle \, + \, \gamma\, |\phi^{\text{\tiny
I}}_{2}(t)\rangle + \ldots,
\end{equation}
while the functional derivative acts on $|\phi^{\text{\tiny
I}}(t)\rangle$ as follows:
\begin{equation}
\frac{\delta}{\delta w_{j}(s)} |\phi^{\text{\tiny I}}(t)\rangle \;
= \, \sqrt{\gamma}\, \frac{\delta}{\delta w_{j}(s)}
|\phi^{\text{\tiny I}}_{1}(t)\rangle \, + \, \gamma\,
\frac{\delta}{\delta w_{j}(s)} |\phi^{\text{\tiny
I}}_{2}(t)\rangle + \ldots;
\end{equation}
the second term must be of order $\gamma$, since the functional
derivative brings down a term proportional to $\sqrt{\gamma}$; for
the same reason, the third term is of order $\gamma^{3/2}$. This
means that the perturbation expansion can be explicitly carried
out, despite the functional derivative appearing in
Eq.~\eqref{eq:colint}, and we get the following results to order
$\gamma$:
\begin{eqnarray}
\text{order $0$:} \;\;\;\;\; \frac{d}{dt}|\phi^{\text{\tiny I}}_{0}(t)\rangle & = & 0, \\
\text{order $\sqrt{\gamma}$:} \;\; \frac{d}{dt}|\phi^{\text{\tiny
I}}_{1}(t)\rangle & = & \xi \sum_{i=1}^{N} A_{i}(t) w_{i}(t)
|\phi^{\text{\tiny
I}}_{0}\rangle, \\
\text{order $\gamma$:} \;\;\;\;\; \frac{d}{dt}|\phi^{\text{\tiny
I}}_{2}(t)\rangle & = & \xi \sum_{i=1}^{N} A_{i}(t) w_{i}(t)
|\phi^{\text{\tiny I}}_{1}(t)\rangle - 2 \xi_{\text{\tiny R}}\xi
\sum_{i,j=1}^{N} \int_{0}^{t} ds A_{i}(t) A_{j}(s) D_{ij}(t,s)
|\phi^{\text{\tiny I}}_{0}\rangle.
\end{eqnarray}
Going back to the Schr\"odinger picture, Eq.~\eqref{eq:col}, to
order $\gamma$, reduces to:
\begin{eqnarray}\label{eq:firstphi}
\frac{d}{dt}|\phi(t)\rangle &= &- \frac{i}{\hbar} H
|\phi(t)\rangle  + \left[ \! \sqrt{\gamma} \;\xi \sum_{i=1}^{N}
A_{i} w_{i}(t)\right.\cr & + & \left. \gamma \xi\!\!
\sum_{i,j=1}^{N} \int_{0}^{t} \!\! ds\, A_{i} A_{j}(s\!-\! t) (
\xi w_{i}(t) w_{j}(s) - 2 \xi_{\text{\tiny R}} D_{ij}(t,s))
\!\right]\! |\phi_0(t)\rangle,
\end{eqnarray}
with $|\phi(t)\rangle $ the total wave function, and
$|\phi_0(t)\rangle $ its zeroth order part.  By use of Eqs.
(30)-(32), this can also be written entirely in terms of
$|\phi(t)\rangle $ as
\begin{equation}\label{eq:secondphi}
\frac{d}{dt}|\phi(t)\rangle =
 \left[ -
\frac{i}{\hbar} H  + \! \sqrt{\gamma} \;\xi \sum_{i=1}^{N} A_{i}
w_{i}(t) -2\gamma \xi \xi_{\text{\tiny R}}\!\! \sum_{i,j=1}^{N}
\int_{0}^{t} \!\! ds\, A_{i} A_{j}(s\!- t) D_{ij}(t,s) \!\right]\!
|\phi(t)\rangle\;.
\end{equation}
We can now see more clearly the non-Markovian nature of the
evolution, because of the presence of the term $A_{j}(s-t)$, which
depends on the past effect of $H$ on $A_{j}$.  When $H=0$, one
immediately sees that by using Eq.~\eqref{eq:fdeff}, the
$|\phi(t)\rangle$ evolution equation of  Eq.~\eqref{eq:secondphi}
reduces to the evolution equation given in Eq.~\eqref{eq:colcom}.

The corresponding equation for the normalized vector
$|\psi(t)\rangle$ defined by Eq.~\eqref{eq:cm3} can now be
obtained by a straightforward application of the chain rule for
differentiation, with the result
\begin{equation}\label{eq:psieq}
\frac{d}{dt}|\psi(t)\rangle=\left[-\frac{i}{\hbar} (H+i\hbar\gamma
O_{ASA}) + \sqrt{\gamma}\sum_{i=1}^N (\xi A_i -\xi_{\text{\tiny
R}} \langle A_i\rangle_t)w_i(t) + \gamma (O_{SA}-\langle
O_{SA}\rangle_t) \right] |\psi(t)\rangle,
\end{equation}
again with the expectations $\langle ...\rangle_{t}$ computed in the
state $|\psi(t)\rangle$.  Here $O_{ASA}$ and $O_{SA}$ are
respectively the anti-self-adjoint and self-adjoint parts of the
operator $O$ defined by the final term inside the square brackets of
Eq.~\eqref{eq:secondphi}, and are given explicitly by
\begin{eqnarray}\label{eq:odefs}
O_{ASA}= - \sum_{i,j=1}^N \int_0^t ds \big( \xi_{\text{\tiny R}}^2
[A_i,A_j(s-t)] + i \xi_{\text{\tiny I}} \xi_{\text{\tiny R}}
\{A_i,A_j(s-t)\} \big) D_{ij}(t,s)\\
O_{SA}=- \sum_{i,j=1}^N \int_0^t ds \big( \xi_{\text{\tiny R}}^2
\{A_i,A_j(s-t)\} + i \xi_{\text{\tiny I}} \xi_{\text{\tiny
R}}[A_i,A_j(s-t)] \big) D_{ij}(t,s).
\end{eqnarray}

The equation for the statistical operator can now be computed by
resorting to relations~\eqref{eq:cm5} and~\eqref{eq:cm6}.  The
calculation proceeds most directly from the $|\phi(t)\rangle$
evolution equation given in Eq.~\eqref{eq:firstphi}, since in this
equation all dependence on the noise is explicit.  To the order to
which we are working, we can replace the $\gamma$ term in this
equation by its expectation ${\mathbb E}_{\mathbb Q} $, giving the
simplified evolution equation
\begin{equation}\label{eq:thirdphi}
\frac{d}{dt}|\phi(t)\rangle = - \frac{i}{\hbar} H |\phi(t)\rangle +
\left[ \! \sqrt{\gamma} \;\xi \sum_{i=1}^{N} A_{i} w_{i}(t)- \gamma
|\xi|^2\!\! \sum_{i,j=1}^{N} \int_{0}^{t} \!\! ds\, A_{i}
A_{j}(s\!-\! t) D_{ij}(t,s) \!\right]\! |\phi_0(t)\rangle.
\end{equation}
It is then straightforward to compute ${\mathbb E}_{\mathbb Q}
[|\phi(t)\rangle \langle \phi(t)|]$, with the result\footnote
{P. Pearle has pointed out that, in the real noise ($\xi=1$) case,
 this equation follows from differentiation of the integrated expression given in Eq.~(4.11) of his {\it Physical
Review} article cited in ~\cite{col}, which he suggests is exact when
time-ordering is included.}

\begin{eqnarray}\label{eq:firstfin}
\frac{d}{dt}\, \rho(t) & = & - \frac{i}{\hbar} [H, \rho(t) ] +
|\xi|^2 \gamma \sum_{i,j=1}^{N} \int_{0}^{t} ds\, D_{ij}(t,s) [
A_{i} \rho(t) A_{j}(s-t) + A_{j}(s-t) \rho(t) A_{i} \nonumber \\
& & \qquad\qquad\qquad\qquad\qquad\qquad\qquad\qquad - A_{i}
A_{j}(s-t) \rho(t) - \rho(t) A_{j}(s-t) A_{i} ].
\end{eqnarray}
As we can see, the above equation, which is correct to order
$\gamma$, depends only on the absolute value of $\xi$; this means
that all physical predictions are, at least to order $\gamma$,
independent of the phase of $\xi$, precisely as in the white-noise
case. As a consequence, in order to compute physical predictions,
one can resort to Eq.~\eqref{eq:col} with $\xi = i$, which is much
simpler since the integro-differential term vanishes, and one is
left with a standard Schr\"odinger equation with a random
hermitian potential.

\section{An alternative construction of the nonlinear collapse equation}
\label{sec:three} We give in this section an alternative
construction of the nonlinear collapse equation in the case of
non-white Gaussian noise.  Instead of starting from a linear
equation and using a change of measure, we work throughout with a
norm-preserving nonlinear equation, and use perturbation theory to
determine the structure of the order $\gamma$ term which
guarantees that the corresponding density matrix evolution will
have the Lindblad form ~\cite{quest} in the Markovian limit.

Thus, returning to Eq.~\eqref{eq:cm1}, we now start from
\begin{equation} \label{eq:nwstr}
\frac{d|\psi(t)\rangle}{dt} = \left[ - \frac{i}{\hbar} H  +
\sqrt{\gamma} \sum_{i = 1}^{N} (\xi A_{i} - \xi_{\text{\tiny R}}
\langle A_{i} \rangle_{t})w_{i}(t)  + \gamma(B_{SA}-\langle
B_{SA}\rangle_t) +\gamma B_{ASA}  \right]
|\psi(t)\rangle,\\
\end{equation}
with $B_{SA}$ and $B_{ASA}$ respectively a self-adjoint operator
and an anti-self-adjoint operator to be determined, and with
$w_i(t)$ now a non-white Gaussian noise obeying
\begin{equation}\label{eq:nwav}
 {\mathbb E}_{\mathbb R} [ w_{i}(t) ] = 0, \qquad {\mathbb
E}_{\mathbb R} [ w_{i}(t_{1}) w_{j}(t_{2}) ] = D_{ij}(t_{1},t_{2})
\end{equation}
with respect to the measure ${\mathbb R}$. We shall determine $B_{SA}$
and $B_{ASA}$ to
simplify the evolution equation for $\rho(t) ={ \mathbb E}_{\mathbb
R}[\hat \rho(t)]= {\mathbb E}_{\mathbb R} [|\psi(t) \rangle \langle
\psi(t)| ]$, in such a way that in the Markovian limit it reduces to
a Lindblad evolution in the first standard form,
\begin{align}\label{eq:stdform}
\frac{d\rho(t)}{dt}=&{\cal L}\rho(t)\cr =&-\frac{i}{\hbar}
[H,\rho(t)] + \sum_{i,j} a_{ij} \left( F_i\rho(t)F_j
-\frac{1}{2}\left\{ F_j^{\dagger}F_i,\rho(t)\right\}\right)~~~,
\end{align}
with $F_i$  suitable functions of $\{A_i\}$, and with the
coefficients $a_{ij}$ determined by the noise expectation
$D_{ij}$.

By construction, Eq. \eqref{eq:nwstr} preserves the normalization of
the state vector $|\psi(t)\rangle$ under time evolution.  From this
equation, and its adjoint, one easily finds that the pure state
density matrix $\hat\rho(t)$ obeys the evolution equation
\begin{align}\label{eq:purestate}
\frac{d\hat\rho(t)}{dt}=&-\frac{i}{\hbar}[H,\hat\rho(t)] +
[\sqrt{\gamma}i\xi_{\text{\tiny I}}\sum_{i=1}^N A_i w_i(t) +\gamma
B_{ASA},\hat \rho(t)] \cr +&\{ \sqrt{\gamma}\xi_{\text{\tiny
R}}\sum_{i=1}^N(A_i-\langle A_i\rangle_t)w_i(t) +\gamma
(B_{SA}-\langle B_{SA}\rangle_t) , \hat\rho(t)\}.
\end{align}
Taking the expectation of this, and retaining terms through order
$\gamma$ but dropping terms of order $\gamma^{3/2}$, we get
\begin{align}\label{eq:avdens}
\frac{d\rho(t)}{dt}&=-\frac{i}{\hbar}[H,\rho(t)]
+\sqrt{\gamma}i\xi_{\text{\tiny I}}
 \sum_{i=1}^N \left[A_i,{\mathbb E}_{\mathbb R} [\hat \rho(t) w_i(t)]
  \right]+\gamma
[B_{ASA},\rho(t)]+\sqrt{\gamma}\xi_{\text{\tiny R}} \sum_{i=1}^N
\left\{A_i,{\mathbb E}_{\mathbb  R}[\hat \rho(t)
w_i(t)]\right\}\cr& -2 \sqrt{\gamma} \xi_{\text{\tiny R}}
\sum_{i=1}^N{\mathbb
  E}_{\mathbb  R}[\hat \rho(t) \langle A_i \rangle_t w_i(t)] +\gamma
\left\{ B_{SA}-\langle B_{SA}\rangle_t, \rho(t) \right\}.
\end{align}

We now use the Furutsu-Novikov formula
\begin{equation}\label{eq:fnnew}
{\mathbb E}_{\mathbb R}[ F[\{w(t)\}]\,w_{i}(t) ] \quad = \quad
\sum_{j=1}^{N} \int_0^{t} ds \, D_{ij}(t,s) {\mathbb E}_{\mathbb
R}  \left[ \frac{\delta F[\{w(t)\}]}{\delta w_{j}(s)} \right]~~~,
\end{equation}
first with $F=\hat \rho(t)$, and then with $F=\langle A_i\rangle_t
\hat \rho(t) = \hat \rho(t) {\rm Tr} \hat \rho(t) A_i$.  By the
chain rule,
\begin{equation}\label{eq:chrule}
\frac{ \delta \hat \rho(t) {\rm Tr} \hat \rho(t) A_i}{\delta
w_j(s)}=\frac{\delta \hat \rho(t)}{\delta w_j(s)} {\rm Tr} \hat
\rho(t) A_i + \hat \rho(t) {\rm Tr}\frac{\delta \hat
\rho(t)}{\delta w_j(s)} A_i,
\end{equation}
so for both choices of $F$ in Eq.~\eqref{eq:fnnew} what we need is
$\delta \hat \rho(t)/\delta w_j(s) $, calculated through terms of
order $\sqrt{\gamma}$.  This can be calculated directly by
integrating the differential equation of Eq.~\eqref{eq:nwstr}. In
terms of the interaction picture operators
\begin{equation}\label{eq:intop}
A_j(s-t)=e^{\frac{i}{\hbar}H(s-t)} A_j e^{-\frac{i}{\hbar}H(s-t)},
\end{equation}
a simple calculation gives
\begin{equation}\label{eq:varderiv}
\frac{\delta \hat \rho(t)}{\delta w_j(s)} =\sqrt{\gamma}
\big[i\xi_{\text{\tiny I}}[A_j(s-t),\hat\rho(t)]+\xi_{\text{\tiny
R}}\{A_j(s-t)-\langle A_j\rangle_s, \hat\rho(t) \}\big].
\end{equation}
Substituting this into Eq.~\eqref{eq:avdens} gives a lengthy
expression, which on algebraic simplification, and noting that
$\langle A_j(s-t) \rangle_t=\langle A_j \rangle_s$, gives a result
that may be summarized as follows.  Let us abbreviate
\begin{equation}\label{eq:abbr}
S_{ij}\equiv \sum_{i,j=1}^N \int_0^t ds D_{ij}(t,s),
\end{equation}
 so that $S_{ij}$ acting on a function of $t,s$
gives a function only of $t$.  Then we find
\begin{align}\label{eq:result}
\frac{d\rho(t)}{dt}=&-\frac{i}{\hbar}[H,\rho(t)] \cr
+&\gamma(\xi_{\text{\tiny R}}^2+\xi_{\text{\tiny I}}^2)S_{ij}
[A_i\rho(t)A_j(s-t)+A_j(s-t)\rho(t)A_i
-A_iA_j(s-t)\rho(t)-\rho(t)A_j(s-t)A_i]\cr +&\gamma[
S_{ij}(\xi_{\text{\tiny R}}^2F_{ij}+i\xi_{\text{\tiny
I}}\xi_{\text{\tiny R}}C_{ij})+B_{ASA},\rho(t)] \cr +&\gamma\{
S_{ij}[\xi_{\text{\tiny R}}^2(C_{ij}-\langle C_{ij} \rangle_t)
+i\xi_{\text{\tiny I}}\xi_{\text{\tiny R}} (F_{ij}-\langle
F_{ij}\rangle_t)] +B_{SA}-\langle B_{SA} \rangle_t,\rho(t) \},
\end{align}
where we have introduced the condensed notations
\begin{align}\label{eq:condnot}
F_{ij}=&[A_i,A_j(s-t)],\cr C_{ij}=&\{A_i,A_j(s-t)\} -2A_i\langle
A_j(s-t)\rangle_t-2\langle A_i\rangle_t A_j(s-t).
\end{align}

Let us now make the following choice of the previously undetermined
operators $B_{SA}$ and $B_{ASA}$,
\begin{align}\label{eq:bvalues}
B_{SA}=&-S_{ij} (\xi_{\text{\tiny R}}^2 C_{ij}+ i\xi_{\text{\tiny
I}}\xi_{\text{\tiny R} } F_{ij}),\cr B_{ASA}=&-
S_{ij}(\xi_{\text{\tiny R}}^2 F_{ij}+i\xi_{\text{\tiny
I}}\xi_{\text{\tiny R}}C_{ij}).
\end{align}
Then the final two lines of Eq.~\eqref{eq:result} cancel to zero,
and we are left with
\begin{align}\label{eq:secfinal}
\frac{d\rho(t)}{dt}=&-\frac{i}{\hbar} [H,\rho(t)] + \gamma
(\xi_{\text{\tiny R}}^2+\xi_{\text{\tiny
I}}^2)\sum_{i,j=1}^N\int_0^t ds D_{ij}(t,s)\cr
\times&[A_i\rho(t)A_j(s-t)+A_j(s-t)\rho(t)A_i
-A_iA_j(s-t)\rho(t)-\rho(t)A_j(s-t)A_i].
\end{align}
Thus, we recover from this approach the evolution of
Eq.~\eqref{eq:firstfin} of Sec. II. In the Markovian limit in
which $D_{ij}(s,t)$ decays rapidly when $s$ is not close to $t$,
we can approximate $s-t\simeq 0$ in Eq.~ \eqref{eq:secfinal}.  We
then have $A_j(s-t)\simeq A_j$, which by assumption commutes with
$A_i$, and the evolution of Eq.~\eqref{eq:secfinal} takes the
first standard form of a Lindblad evolution,
\begin{align}\label{eq:lind1}
\frac{d\rho(t)}{dt}=&-\frac{i}{\hbar} [H,\rho(t)] + \gamma
(\xi_{\text{\tiny R}}^2+\xi_{\text{\tiny
I}}^2)\sum_{i,j=1}^N\int_0^t ds D_{ij}(t,s)\cr
\times&[A_i\rho(t)A_j+A_j\rho(t)A_i -\{A_iA_j,\rho(t)\}].
\end{align}
Eq.~\eqref{eq:lind1} also follows, without the Markovian
approximation, if one neglects the Hamiltonian $H$, since then the
interaction and Schr\"odinger pictures coincide, and one has
exactly $A_j(s-t)=A_j$.

\section{Change of Measure and Other Concluding remarks}
\label{sec:four}
\subsection{Change of Measure}

In Secs. II and III we have given two different derivations  of
the time evolution equation for the normalized wave function
$|\psi(t)\rangle$, and  of the corresponding evolution equation
for the density matrix $\rho(t)$.  By construction, the density
matrix evolutions of Eq.~\eqref{eq:firstfin} and
Eq.~\eqref{eq:secfinal} are the same. However, a comparison of the
operators $O_{SA}$ and $O_{ASA}$ of Sec. II with the operators
$B_{SA}$ and $B_{ASA}$ of Sec. III shows that these are not the
same, and hence the corresponding evolution equations
 for $|\psi(t)\rangle$ of Eq.~\eqref{eq:psieq}
 and Eq.~\eqref{eq:nwstr} are not the same. This
means that our two constructions correspond to inequivalent
unravelings of the same density matrix evolution: in general the
noises, expectations, and wave functions $|\psi(t)\rangle$ of Sec.
II and Sec. III are not the same, even though they lead to the
same evolution equation for the noise-averaged density matrix.

However, in certain special cases the formulations of Secs. ~II
and ~III are related by a time-dependent shift in the mean values
of the Gaussian noise variables.  In order for the functions
$|\psi(t)\rangle$ used in the two derivations to be identical,
they must obey the same time evolution equation. Changing
notation, by denoting the noise of Sec. III as $\tilde w_i(t)$, we
find that the evolution equations for $|\psi(t)\rangle$ of
Eq.~\eqref{eq:psieq} and Eq.~\eqref{eq:nwstr} become identical
when the noises $\tilde w_i(t)$ and $w_i(t)$ are related to
leading order in $\sqrt\gamma$ by
\begin{equation}\label{eq:chvar1}
\sum_{i=1}^N \tilde w_i(t) A_i=\sum_{i=1}^N w_i(t) A_i -2
\sqrt\gamma\,\xi_{\text{\tiny R}}\sum_{i,j=1}^N \int_0^t ds
D_{ij}(t,s)[A_i\langle A_j(s-t)\rangle_t+\langle A_i\rangle_t
A_j(s-t)],
\end{equation}
with
\begin{equation}\label{eq:zerocond}
{\mathbb E}_{\mathbb R}[\tilde w_i(t) \tilde w_j(s)] ={\mathbb
E}_{\mathbb Q}[w_i(t) w_j(s)]=D_{ij}(t,s)
\end{equation}
to zeroth order in $\sqrt\gamma$.

 Equation \eqref{eq:chvar1} is
consistent only if the operators $A_j(s-t)$ can be expanded over a
basis of the time zero operators $A_i$, with c-number
coefficients, that is
\begin{equation}\label{eq:expansion}
A_j(s-t)= \sum_{i=1}^N K_{ji}(s-t) A_i.
\end{equation}
This is automatically true (i) when the Hamiltonian $H$ vanishes,
since then $A_j(s-t)$ is time independent, and (ii) in the white
noise limit, since then $D_{ij}(t,s) \propto \delta(t-s)$, and so
Eq.~\eqref{eq:chvar1} only involves $A_j(0)=A_j$. It is
approximately true (iii) whenever a Markovian approximation to the
time evolution is valid. In general, however,
Eq.~\eqref{eq:chvar1} involves on the right hand side operators
$A_j(s-t)$ that are linearly independent of the operators $A_i$,
and so cannot be satisfied.

When Eq.~\eqref{eq:expansion} holds, then Eq.~\eqref{eq:chvar1}
simplifies to take the form
\begin{equation}\label{eq:chvar2}
 \sum_{i=1}^N \tilde w_i(t)
A_i=\sum_{i=1}^N w_i(t) A_i -\sum_{i=1}^N K_i(t) A_i,
\end{equation}
with $K_i(t)$ given by
\begin{align}\label{eq:kdeff}
K_i(t)=& 2\sqrt\gamma\,\xi_{\text{\tiny R}}\sum_{j=1}^N \int_0^t
ds [D_{ij}(t,s)\langle A_j(s-t)\rangle_t + \sum_{m=1}^N
D_{jm}(s-t) \langle A_j \rangle_t K_{mi}(s-t) ].
\end{align}
Eq.~\eqref{eq:chvar2} can clearly be satisfied by making a
c-number shift in the noise variable for each $i$,
\begin{equation}\label{eq:chvar3}
\tilde w_i(t)=w_i(t)- K_i(t)
\end{equation}
The relation at time $t$ between the measure ${\mathbb R} $ of Sec.
III, in which $\tilde w_i(t)$ has zero mean, and the measure
${\mathbb Q}$ of Sec. II, in which $w_i(t)$ has zero mean, is then
to first order in $\sqrt\gamma$, for any argument $O$,
\begin{equation}\label{eq:chmeas}
{\mathbb E}_{\mathbb R}[O]={\mathbb E}_{\mathbb Q}[W(t) O].
\end{equation}
Here the weighting factor $W(t)$ is given by
\begin{equation}\label{eq:wdeff}
W(t)=1+\sum_{i=1}^N \int_0^t ds C_i(t,s) w_i(s),
\end{equation}
where $C_i(t,s)$ obeys the integral equation
\begin{equation}\label{eq:ceq}
K_i(t)=\sum_{j=1}^N\int_0^t ds D_{ij}(t,s)C_j(t,s) .
\end{equation}
This choice of $C_i(t,s)$ guarantees that when $O$ is taken as
$\tilde w_i$ in Eq.~\eqref{eq:chmeas}, one finds that ${\mathbb
E}_{\mathbb R} [\tilde w_i]=0$, as needed. An analogous c-number
shift of the noise variable enters in comparing Eq.~\eqref{eq:str}
of Sec. I with the white noise limit of Eq.~\eqref{eq:colcomcon} in
Sec. IIA.

\subsection{Final Remarks}

 In the preceding sections we have shown that the white noise
formalism is robust under a generalization to the physically more
realistic assumption of non-white noise.  Both the proof of state
vector reduction, and the ``imaginary time'' trick for calculating
physical effects of the noise, carry over to the non-white noise
case, to leading quadratic order in the noise strength.

We wish here to elaborate on some implications of our calculations
for models of the non-white noise.  We recall that the noise
autocorrelation  $D_{ij}(t_1,t_2)$ is defined as the expectation
${\mathbb E}_{\mathbb Q} [w_i(t_1)w_j(t_2)]$, and that the
condition for state vector reduction is that the time integral
\begin{equation}\label{eq:fdefs}
F_{ij}(t)=\int_0^t ds D_{ij}(t,s)
\end{equation}
should be a positive definite matrix in the limit as $t\to
\infty$.\footnote{ More generally, as can be seen from
Eq.~\eqref{eq:redu},
 the reduction requirement is satisfied when $\int_0^t ds
 F_{ij}(s) \to \infty$ as $t \to \infty$.}  Let us now assume time translation invariance, which
implies that $D_{ij}(t,s)=D_{ij}(t-s)$, and  investigate what the
requirement for state vector reduction means in terms of the
spectral decomposition of $D_{ij}$.  Writing
\begin{equation}\label{eq:spectral}
D_{ij}(t-s)=\int_0^{\infty} d\omega \gamma_{ij}(\omega)
\cos\omega(t-s)
\end{equation}
we have
\begin{equation}\label{eq:fint}
F_{ij}(t)=\int_0^{\infty} d\omega \gamma_{ij}(\omega)
\frac{\sin\omega t}{\omega} =\int_0^{\infty} \frac{du}{u} \sin u
\gamma_{ij}(u/t).
\end{equation}
Thus, assuming that $\gamma(\omega)$ is smooth in the neighborhood
of $\omega=0$, we find
\begin{equation}\label{eq:limit}
 \lim_{t\to\infty}F_{ij}(t)= \gamma_{ij}(0)\int_0^{\infty} \frac{du}{u} \sin u =
\gamma_{ij}(0) \pi.
\end{equation}
Hence, the reduction requirement is satisfied when
$\gamma_{ij}(0)$ is a positive definite matrix in $i,j$. In
particular, the spectral weight $\gamma_{ij}(\omega)$ can have a
cutoff at a finite upper limit $\omega=\omega_{\rm max}$, without
in any way affecting the reduction argument.  The possibility of
such an upper cutoff has been discussed in the review of Bassi and
Ghirardi ~\cite{cm}, and as noted by  Adler and Ramazanoglu
~\cite{adran}, is suggested on physical grounds by existing upper
limits on noise-induced gamma ray emission.

Let us next consider the rate of secular energy increase induced
by the noise, taking advantage of the ``imaginary noise'' trick to
write the noise term as an addition to the Hamiltonian.  We
consider the simple model describing a particle of mass $m$ moving
in one dimension, with a non-white noise coupling to its
coordinate $x$, with Hamiltonian
\begin{equation}\label{eq:1dimham}
H=\frac{p^2}{2m}-C w_t x.
\end{equation}
Because the noise term does not commute with $p$, the kinetic
energy $p^2/(2m)$ increases over time.  We have for the expected
rate of energy gain,
\begin{equation}\label{eq:engain}
\frac {d}{dt}  {\mathbb E}\left[ \frac {p^2}{2m} \right] = m^{-1}
 {\mathbb E} \left[ p \frac{dp}{dt} \right].
 \end{equation}
{}From the Heisenberg equations of motion implied by
Eq.~\eqref{eq:1dimham} we find $dx/dt=p,~dp/dt=Cw_t$, and so
\begin{equation}
p(t)=p(0)+C\int_0^t du w_u
\end{equation}
Hence the rate of energy gain is
\begin{equation}
\frac {d}{dt}  {\mathbb E}\left[ \frac {p^2}{2m} \right] =
\frac{C^2}{m} \int_0^t du
 {\mathbb E} \left[ w_t w_u \right]=\frac{C^2}{m} \int_0^t du
 D(t,u) = \frac{C^2}{m}F(t),
\end{equation}
and therefore is governed by the same integral over the
autocorrelation function as the reduction rate.  Thus, when the
reduction condition $\lim_{t\to \infty} F(t) > 0$ is obeyed, the
rate of noise-induced energy production is nonzero at large times,
a result which readily generalizes to more realistic reduction
models. As  reviewed in ~\cite{rev}, this leads to various upper
bounds on the noise strength.  This conclusion can be evaded in
the generic case of multiple operators $A_i$, if the only
nonvanishing eigenvalue of $F_{ij}(t)$ as $t \to \infty$ is the
one associated with the total energy. An interesting model where
this case is realized, but in which localizing reduction still
occurs in an approximate sense, is given by taking $D_{ij}(t,s)$
to be a correlation function associated with thermal noise, as
might be expected if state vector reduction is induced by some
type of cosmological relic field.  A detailed examination of the
thermal noise model will be given elsewhere \cite{adlerbassi}.

\section{Acknowledgments}
The work of SLA was supported in part by the Department of Energy
under grant no DE-FG02-90ER40542, and part of the work was done
while he was at the Aspen Center for Physics.  The work of AB was
supported in part by DGF (Germany) and by the EU grant
ERG\;044941-STOCH-EQ; part of his work was done while visiting the
Institute for Advanced Study in Princeton, which he wishes to
thank for its support, and Clare Hall in Cambridge.  We wish to
thank Philip Pearle for helpful correspondence.

\end{document}